\documentclass[12pt]{article}

\usepackage[left=2.8cm,right=2.8cm,top=2.9cm,bottom=2.9cm]{geometry}
\usepackage[french,english]{babel}
\usepackage[colorlinks=true,linkcolor=black,citecolor=blue,urlcolor=blue,filecolor=black]{hyperref}

\usepackage[utf8]{inputenc}
\usepackage[T1]{fontenc}
\usepackage{latexsym}
\usepackage{amsmath}
\usepackage{slashed}
\usepackage{amsfonts}
\usepackage{verbatim}
\usepackage{mathrsfs}
\usepackage{dsfont}

\newcommand{\p}{\partial}

\newcommand{\be}{\begin{eqnarray}}
\newcommand{\en}{\end{eqnarray}}
\newcommand{\badat}{\begin{alignedat}}
\newcommand{\eadat}{\end{alignedat}}
\newcommand{\bitm}{\begin{itemize}}
\newcommand{\eitm}{\end{itemize}}
\newcommand{\bmat}{\begin{pmatrix}}
\newcommand{\emat}{\end{pmatrix}}
\newcommand{\ba}{\begin{align}}
\newcommand{\bas}{\begin{align*}}
\newcommand{\ab}{\end{align}}
\newcommand{\bse}{\begin{subequations}}
\newcommand{\ese}{\end{subequations}}
\newcommand{\gt}{\rightarrow}

\def\cJ{\mathcal{J}}

\def\m{\mu}
\def\n{\nu}

\def\cJ{{\cal J}}

\def\cL{{\cal L}}

\def\cT{{\cal T}}

\def\cY{{\cal Y}}

\def\bea{\begin{eqnarray}}
\def\eea{\end{eqnarray}}
\def\ba{\begin{array}}
\def\ea{\end{array}}
\def\bec{\begin{center}}
\def\ec{\end{center}}
\def\ba{\begin{align}}
\def\ena{\end{align}}

\def\12{\frac{1}{2}}

\def\pr{\partial}

\begin{document}

\title{\textbf{Extended symmetries at the black hole horizon}}

\author{Laura Donnay$^{1}$, Gaston Giribet$^{1,2}$, Hern\'an A. Gonz\'alez$^{1}$, Miguel Pino$^{3}$}
\date{}
\maketitle

\begin{center}

\smallskip
\smallskip
{$^1$ Universit\'{e} Libre de Bruxelles and International Solvay Institutes}\\
{{\it ULB-Campus Plaine CPO231, B-1050 Brussels, Belgium.}}

\smallskip
\smallskip

{$^2$ Departamento de F\'{i}sica, Universidad de Buenos Aires and IFIBA-CONICET}\\
{{\it Ciudad Universitaria, Pabell\'{o}n I, 1428, Buenos Aires, Argentina.}}

\smallskip
\smallskip

{$^3$ Departamento de F\'{i}sica, Universidad de Santiago de Chile}\\
{{\it Avenida Ecuador 3493, Estaci\'{o}n Central, 9170124, Santiago, Chile.}}




\end{center}

\bigskip

\begin{abstract}
We prove that non-extremal black holes in four-dimensional general relativity exhibit an infinite-dimensional symmetry in their near horizon region. By prescribing a physically sensible set of boundary conditions at the horizon, we derive the algebra of asymptotic Killing vectors, which is shown to be infinite-dimensional and includes, in particular, two sets of supertranslations and two mutually commuting copies of the Virasoro algebra. We define the surface charges associated to the asymptotic diffeomorphisms that preserve the boundary conditions and discuss the subtleties of this definition, such as the integrability conditions and the correct definition of the Dirac brackets. When evaluated on the stationary solutions, the only non-vanishing charges are the zero-modes. One of them reproduces the Bekenstein-Hawking entropy of Kerr black holes. We also study the extremal limit, recovering the NHEK geometry. In this singular case, where the algebra of charges and the integrability conditions get modified, we find that the computation of the zero-modes correctly reproduces the black hole entropy. Furthermore, we analyze the case of three spacetime dimensions, in which the integrability conditions notably simplify and the field equations can be solved analytically to produce a family of exact solutions that realize the boundary conditions explicitly. We examine other features, such as the form of the algebra in the extremal limit and the relation to other works in the literature.
\end{abstract}

\newpage

\section{Introduction}

There has been a renewed interest in the study of infinite-dimensional symmetries in the near horizon region of non-extremal black holes. This is mainly motivated by the proposal in \cite{Hawking, HPS}. There, it was argued that the conserved charges associated to a particular symmetry of this kind, known as supertranslation \cite{BMS1, BMS2, BMS3} could lead to an ingenious way of circumventing no-hair theorems and eventually solve the information black hole paradox \cite{Hawking:1976ra}. This idea has attracted remarkable attention recently \cite{Donnay:2015abr, Blau:2015nee, Afshar:2015wjm, Averin:2016ybl, Afshar:2016wfy, Setare:2016jba, Mao:2016pwq, Averin:2016hhm, Afshar:2016uax, Todos} and also raised some controversy \cite{Porrati}. 

Infinite-dimensional symmetries in the near black horizon region have already been discussed in the literature \cite{Carlip1, Koga, Hotta:2002mq, Hotta:2000gx} and they have been studied in relation to different contexts, such as the membrane paradigm \cite{Penna:2015gza, Eling:2016xlx}. In the recent paper \cite{Donnay:2015abr}, the supertranslation symmetry appearing close to the black hole horizon was explicitly worked out by studying the asymptotic behavior of the metric excitations close to the horizon. It has been shown there that, for a suitable choice of boundary conditions, the infinite-dimensional local symmetries at the horizon get enhanced and, in addition to supertranslation, the algebra of charges include the Virasoro algebra. Here, aimed at further investigating this phenomenon, we will extend the analysis of \cite{Donnay:2015abr}. We will show that, by considering a more general set of boundary conditions, which in particular admits dependence of time, the asymptotic isometries in the near horizon region of non-extremal black holes are further enhanced in such a way that a new set of supertranslations appears. We will explicitly work out the asymptotic symmetry group together with the charge algebra of the extended symmetries both in three and four spacetime dimensions. In four dimensions, we will discuss the subtleties in the definition of those charges, such as the additional conditions coming from demanding integrability. In three dimensions, we will present a family of exact solutions that explicitly realize the proposed boundary conditions at the horizon. We will also discuss the limit in which the black hole becomes extremal and  perform a canonical analysis of the charges in this case.

The paper is organized as follows: In section 2, we discuss the case of black hole horizons in four spacetime dimensions. First, we define the boundary conditions in the near horizon limit, which in particular allow for time-dependent configurations. We derive the asymptotic Killing vectors that preserve such conditions and show they span an infinite-dimensional algebra. Then, we study the conserved charges associated to such asymptotic symmetries, together with the additional conditions coming from imposing the field equations, and the further constraints coming from integrability. We derive the algebra of charges, which is seen to be infinite-dimensional as well. We discuss the evaluation of the charges on solutions whose physical interpretation is under control; we consider the case of stationary black holes and  Rindler horizons and we show that the zero modes of the charges gather in particular the black hole entropy. We will also analyze the extremal limit, where we obtain the NHEK geometry and we obtain similar results as in the previous case. In section 3, we discuss the three-dimensional case, where the Einstein field equations can be solved exactly. This provides a family of solutions parameterized by three arbitrary functions that realizes the boundary conditions explicitly. We conclude in section 4 with a discussion of open questions.

\section{Four-dimensional horizons}

\subsection{Boundary conditions} \label{EstaEEE}

We will be concerned with the behavior of the gravitational field near the black hole horizon. Therefore, we will begin our analysis in this subsection by discussing the form of the metric close to smooth null codimension-one surfaces in four dimensional spacetimes. This will allow us to define physically sensible boundary conditions on the black hole horizon. We will do so by prescribing an asymptotic expansion around the null surface.

Following \cite{Tamburino:1966zz}, let us consider Eddington-Filkenstein coordinates, which in particular include the advanced time coordinate $v$ such that  a null surface is defined by
\begin{equation}
g^{\mu \nu} \pr_{\mu}v  \pr_{\nu}v=0.
\end{equation}
We define a ray as the vector tangent to this surface, $k^{\alpha}=g^{\alpha \beta}\pr_{\beta}v$. In a null frame, we consider 
the temporal coordinate $x^0=v$, and $x^1=\rho$ as the affine parameter of the generator $k^{\mu}$ such that $k^{\mu}=\frac{dx^{\mu}}{d\rho }=\delta^{\mu}_\rho$. Other two coordinates $x^A$ ($A=2,3$) are chosen as parameters constant along each ray, $k{^\nu}\partial_{\nu}{x^A}=0$. 
In terms of the metric, the former impositions are translated into the algebraic conditions
\begin{equation}
 \label{eql:null3}
g^{vv}=0,\, g^{\rho v}=1, \, g^{vA}=0,
\end{equation}
that is
\begin{equation}
 \label{eql:null333}
g_{\rho \rho}=0, \, g_{v\rho}=1, \, g_{\rho A}=0.
\end{equation}
Let us set a null surface at $\rho=0$. Assuming this is a non-expanding surface, the remaining components of the metric close to this region behave like \cite{Booth:2012xm}
\begin{equation}
\begin{split}
 \label{eql:null4}
g_{v v}&=-2 \rho \kappa +O(\rho^2),\\ 
g_{v A}&= \rho \theta_A +O(\rho^2),\\
g_{AB}&=\Omega_{AB}+\rho \lambda_{AB}+O(\rho^2),
\end{split}
\end{equation}
where functions $\kappa$, $\theta_A$, $\Omega_{AB}$, and $\lambda_{AB} $ in principle depend on the coordinates $x^A$ and $v$, with $\Omega_{AB}$ being assumed to be non-degenerate. The boundary conditions on null surfaces we consider include, in particular, black hole horizons. However, they also include time-dependent metrics. We will discuss below additional restrictions on the configurations, such as the ones required to describe isolated horizons. 

In other words, we will consider metrics of the form 
\begin{equation}
ds^2=-2 \kappa \rho \ dv^2+2d\rho dv + 2\theta_A\rho \ dvdx^A+(\Omega_{AB}+\lambda_{AB}\rho )dx^Adx^B +\Delta g_{ij}\ dx^{i}dx^{j },\label{formA}
\end{equation}
$\Delta g_{i j}$ being functions of order ${O}(\rho^2)$ ($i,j \in \{v,A\}$). Actually, it is always possible to find a coordinates system in which the metric close to a smooth null surface admits to be written in the form (\ref{formA}) \cite{Moncrief:1983xua, Chrusciel}.

Notice that here we are considering the possibility of function $\kappa $ (which will be ultimately associated to the surface gravity of the horizon) to vary. This generalizes the analysis of \cite{Donnay:2015abr}, where $\kappa$ was assumed to be a fixed constant. Another difference with \cite{Donnay:2015abr} is that we consider here a more general dependence on the variables; for instance, we are not assuming $\theta_A$, $\Omega_{AB}$ to depend only on $x^{A}$, but they in principle can depend on the advanced time as well.

\subsection{Field equations} \label{EEE}

The next step is solving Einstein's field equations in a way consistent with the expansion above. 

It is often convenient to choose an off-diagonal gauge for the $x^A$ part of the metric in such a way that the 2-dimensional induced metric on the horizon at fixed $v$ is written in the conformal form
\begin{equation}
\label{conformalhorizon}
\Omega_{AB} dx^A dx^B=  4\Omega \ \frac{dz d\bar{z} }{(1+z\bar{z})^2} ,
\end{equation}
with the conformal factor $\Omega $ being a function of $z$, $\bar{z}$, and $v$.

Solving Einstein's field equations introduces additional restrictions, which come in the form of relations among the metric functions and their first and second derivatives. For instance, using the gauge (\ref{conformalhorizon}), close to the horizon (i.e. in the limit $\rho\simeq 0$), one finds that the $(v,v)$ component of Einstein's equations at order ${\mathcal O}(\rho^0)$ can be algebraically solved for $\kappa$, yielding 
\begin{equation}
 \label{vv}
\partial_v^2 \Omega = \frac{1}{2} \Omega^{-1}(\partial_v\Omega )^2 +\kappa \partial_v \Omega.
\end{equation}
On the other hand,  components $(v,x^A)$ of the field equations gives relations for $\theta_A$
\begin{equation}
  \label{vxA}
	\partial_v (\theta_A  \Omega) = -\partial_v \partial_A \Omega + \Omega^{-1} \partial_v \Omega \partial_A \Omega-2\Omega \partial_A \kappa.
\end{equation}
Equations (\ref{vv}) and (\ref{vxA}) will be important later for solving the integrability constraints of the conserved charges. It is worthwhile pointing out that, while important for those purposes, Einstein's equations will not be used in the next subsection, where the asymptotic symmetries are derived independent of the dynamics.

\subsection{Asymptotic symmetries}
\label{s4D}
Having defined the boundary conditions \eqref{eql:null333} and \eqref{eql:null4}, the next step is analyzing the residual symmetries respecting such conditions. For the set \eqref{eql:null333}, this amounts to solve the set of equations
\begin{equation}
 \label{eql:liexact}
\cL_{\chi} g_{\rho \rho}=0, \, \cL_{\chi} g_{v\rho}=1, \, \cL_{\chi} g_{\rho A}=0,
\end{equation}
whose solution is given by
\begin{equation}
\label{solexact}
\badat{3}
&\chi^{v} = f, \\
&\chi^{\rho } = Z - \rho \pr_{v}f + \pr_{A}f \int^\rho_0 d\rho' g^{AB}g_{vB},  \\
&\chi^{A} = Y^A -\pr_{B}f \int^\rho_0 d\rho' g^{AB} ,  
\eadat
\end{equation}
where $f$, $Z$ and $Y^A$ are functions that do not depend on $\rho$. Preserving conditions \eqref{eql:null4} demand  $\cL_{\chi} g_{i j}=\delta g_{i j}$. In other words,
\begin{equation}
 \label{eql:lieasymp}
 \badat{3}
&\cL_{\chi} g_{v v}=-2 \rho \delta \kappa+\delta g^{(2)}_{vv}\rho^2+o(\rho^2),\\ 
&\cL_{\chi} g_{vA}= \rho \delta \theta_A + \delta g^{(2)}_{vA} \rho^2+o(\rho^2),\\ 
&\cL_{\chi} g_{AB}=\delta \Omega_{AB}+\rho \delta \lambda_{AB}  + \delta g^{(2)}_{AB} \rho^2+o(\rho^2).
\eadat
\end{equation}

By expanding the asymptotic Killing vectors \eqref{solexact} in powers of $\rho$, one sees that the first equation of \eqref{eql:lieasymp} leads to
\begin{equation}
\delta \kappa = Y^A\partial_A \kappa +\partial_v (\kappa f) + \partial_v^2f-\theta_A\partial_vY^A-g_{vv}^{(2)}Z \label{Guco},
\end{equation}
and
\begin{equation}
\partial_v Z -\kappa \ Z = 0.
\end{equation}
The second equation of \eqref{eql:lieasymp} gives
\begin{equation}
\label{theeq}
\theta_A Z +\partial_v Y^B\Omega_{BA} + \partial_A Z =0,
\end{equation}
and
\begin{equation}
\delta \theta_A =\theta_B \partial_A Y^B + Y^B\partial_B\theta_A + f \partial_v\theta_A -2\kappa \partial_A f -2\partial_v\partial_A f + \Omega^{BD} \partial_v\Omega_{AB} \partial_D f+2g_{vA}^{(2)}Z.
\end{equation}
From the last equation of \eqref{eql:lieasymp} one finds 
\begin{equation}
\delta\Omega_{A B} = Z \lambda_{AB} +f\partial_v\Omega_{AB} + {\mathcal L}_{Y}\Omega_{AB},
\end{equation}
with  ${\mathcal L}_{Y}$ denoting the Lie derivative along the vector field $Y^{A}$.
Moreover, 
\begin{eqnarray}
\delta\lambda_{AB}= 2 Z g_{AB}^{(2)} -\lambda_{AB}\partial_v f +f\partial_v \lambda_{AB} +\cL_Y \lambda_{AB} +\theta_A\partial_B f+\theta_B\partial_A f -2\nabla_A\nabla_B f, \label{Guco17}
\end{eqnarray}
where $\nabla_A$ stands for the covariant derivative with respect to $\Omega_{AB}$. 

In what follows, we will assume that the leading terms of the asymptotic Killing vector $\chi$ does not depend on the fields\footnote{This is usually referred to as the assumption of the boundary conditions to be ``state independent'', what means that the form of the asymptotic Killing vectors are not considered to depend explicitly of the charges.}. From relation \eqref{theeq} this necessarily implies \footnote{The imposition $Z=0$ can also be reached by consistency of the equation of motion \eqref{vv} with the asymptotic Killing equation \eqref{eql:liexact} and \eqref{eql:lieasymp}.}
\begin{equation}
Z=0, \,\,\,\, \partial_v Y^B=0.
\end{equation}
Therefore, the form of the asymptotic Killing vectors preserving boundary conditions  \eqref{eql:null333} and \eqref{eql:null4} is given by
\begin{equation}
\badat{3}
\label{vectores4D}
&\chi^{v} = f(v,x^A), \\
&\chi^{\rho } =-\pr_v f \rho + \frac{1}{2} \Omega^{AB} \theta_A \partial_{B} f \rho^2 + O(\rho^3),  \\
&\chi^{A} = Y^A(x^B) + \Omega ^{AC}\partial_{C}f  \rho + \frac{1}{2} \Omega^{AD} \Omega^{CB} \lambda_{DB} \pr_{C} f \rho^2 + O(\rho^3), 
\eadat
\end{equation}
where $\Omega^{AB}$ is the inverse of $\Omega_{AB}$.

The corresponding variation of the fields read 
\begin{equation}
\badat{3}
&\delta_{\chi} \kappa =Y^{A} \pr_{A} \kappa + \pr_v (\kappa f) + \pr_v^2 f, \label{deltakappa} \\ 
&\delta_{\chi}  \Omega_{AB} = f\partial_v\Omega_{AB} + {\mathcal L}_{Y}\Omega_{AB},\\
&\delta_{\chi} \theta_A =\cL_Y \theta_{A}+ f \partial_v\theta_A -2\kappa \partial_A f -2\partial_v\partial_A f +\Omega^{BD} \partial_v\Omega_{AB} \partial_D f ,\\
&\delta_{\chi}  \lambda_{AB}=f\partial_v \lambda_{AB} -\lambda_{AB}\partial_v f +\cL_Y \lambda_{AB} +\theta_A\partial_B f+\theta_B\partial_A f -2\nabla_A\nabla_B f.
\eadat
\end{equation}

By introducing a modified version of Lie brackets \cite{Barnich3}
\begin{equation}
\label{modified}
  [\chi_1,\chi_2]=\mathcal{L}_{\chi_1}\chi_2-\delta_{\chi_1} \chi_2+\delta_{\chi_2} \chi_1 ,
\end{equation}
which suffices to take into account the dependence of the asymptotic Killing vectors upon the functions in the metric, one finds that the algebra of these vectors closes; namely
\begin{equation}
  [\chi(f_1,Y_1^A),\chi(f_2,Y_2^A)]=\chi(f_{12},Y_{12}^A),
	\label{closure}
\end{equation}
where
\begin{equation} \label{modifieddd}
\begin{split}
f_{12}=f_1 \pr_{v} f_2-f_2\pr_{v} f_1+Y_1^A \pr_{A} f_2-Y_2^A\pr_{A} f_1,\quad Y_{12}^A=Y_1^B  \pr_{B}  Y_2^A -Y_2^B  \pr_{B} Y_1^A.  
\end{split}
\end{equation}

\subsection{Surface charges} \label{conno}

Having obtained the asymptotic Killing vectors (\ref{vectores4D}), the next step is computing the associated charges. The covariant approach \cite{Barnich:2001jy, Barnich:2007bf} permits to define the variation of surface charges as
\begin{multline}
\label{eqA:21z}
\slashed{\delta}Q_{\xi}[g ; h]= \frac{1}{16 \pi G} \int  (d^2 x)_{\mu \nu}  \sqrt{-g} \,\Bigg[ \xi^{\nu} {\nabla}^{\mu} h - \xi^{\nu} {\nabla} _{\sigma} h^{\mu \sigma} +\xi_{\sigma} {\nabla}^{\nu} h^{\mu \sigma}\\+\frac{1}{2}h {\nabla}^{\nu} \xi^{\mu} +\frac{1}{2}h^{\nu \sigma} \left({\nabla}^{\mu} \xi_{\sigma}-{\nabla}_{\sigma} \xi^{\mu}\right)-(\mu \leftrightarrow \nu)\Bigg] .
\end{multline}
where $\xi$ is an asymptotic Killing vector and $h_{\m \n}=\delta g_{\mu \nu}$ corresponds to a variation of the metric $g_{\m \n}$ within the family of solutions; the symbol $\slashed{\delta}$ stands to emphasize that this expression is not necessarily integrable; while $(d^2 x)_{\mu \nu} = (1/4)\varepsilon _{\mu \nu \alpha \beta } dx^{\alpha }\wedge dx^{\beta }$. 

Using the asymptotic Killing vectors \eqref{vectores4D} and our boundary conditions \eqref{eql:null333} and \eqref{eql:null4}, expression \eqref{eqA:21z} evaluated at the horizon is given by
\begin{equation} 
\badat{2}
\slashed{\delta} Q_{[f,Y^A]}=&\frac{1}{16\pi G} \int d^2x \ [ 2f\kappa \delta(\sqrt{\mathrm{det} \Omega}) + 2 \pr_v f \delta( \sqrt{\mathrm{det} \Omega})- 2 f  \pr_{v}\delta(\sqrt{\mathrm{det} \Omega}) \\
&+\frac{1}{2}f \sqrt{\mathrm{det} \Omega} \,(\Omega^{AB}\Omega^{CD}-\Omega^{AC}\Omega^{BD})\p_v \Omega_{CD}\delta \Omega_{AB} - Y^A \delta(\theta_A \sqrt{\mathrm{det} \Omega})].
\label{Qgen}
\eadat
\end{equation}
We will see below that integrability requires further conditions. 

It is worth noticing that in three spacetime dimensions, where the metric on the horizon is characterized by a single function, the term that contains the factor $(\Omega^{AB}\Omega^{CD}-\Omega^{AC}\Omega^{BD})$ vanishes identically. This remark will be of importance later, when we will discuss the restrictions coming from the integrability conditions of the charges. This results in a notable difference between four and three-dimensional cases.

Focusing on the four-dimensional case for the moment, let us choose on the two-dimensional manifold defined by $\rho=0$ at fixed $v$ a coordinates system in which the metric $\Omega_{AB}$ results to be locally, conformally equivalent to the two-sphere. Namely, such as we did in (\ref{conformalhorizon})
\begin{equation}
\label{conformalhorizon2}
\Omega_{AB} = \Omega \gamma_{AB}, \, \ \ \gamma_{AB} dx^A dx^B= \frac{4}{(1+z\bar{z})^2} dz d\bar{z}.
\end{equation}
This implies that the vectors $Y^A$ are conformal Killing vectors on the two-sphere, and this eventually yields two Virasoro algebras \cite{Donnay:2015abr}. With no major modification we could have considered the case where $\Omega_{AB}$ is generic, and therefore $Y^A$ generates the group of diffeomorphisms on the sphere \cite{Penna:2015gza}. In the conformal gauge (\ref{conformalhorizon2}), the charge variation (\ref{Qgen}) renders
\begin{equation}
\slashed{\delta} Q_{[f,Y^A]}=\frac{1}{16\pi G} \int d z d\bar{z} \sqrt{\gamma} \left( 2f\kappa \delta \Omega + 2 \pr_v f \delta \Omega - 2 f \delta \pr_{v} \Omega+f \frac{\p_v \Omega}{\Omega} \delta \Omega  - Y^A \delta(\theta_A \Omega)  \right).
\label{QLaura}
\end{equation}

As mentioned, this expression is not integrable in general. This is because of two reasons: First, $\kappa $ is in general allowed to vary on the phase space ($\delta \kappa \neq 0$). Secondly, the fourth term  contains a factor $f\Omega^{-1}\partial_v\Omega \delta \Omega$, which spoils integrability because it involves both $\Omega $ and its derivative. The latter term, as we anticipated above, will not be present in the case of three spacetime dimensions. In four dimensions, on the contrary, it demands further restrictions on the configuration space, which we will discuss later in subsection \ref{integrabilitY}. In what follows, we will be mainly concerned with the case 
of fixed, constant $\kappa$.

\subsection{Isolated horizons}

\subsubsection{Fixed temperature configurations}

Let us restrict ourselves to the case of isolated horizons, where we deal with fixed temperature configurations. From the physical point of view, this case is of particular importance as it is the relevant one to describe horizons in (quasi) equilibrium. From the computational point of view, on the other hand, if one assumes that $\kappa$ is constant, then the solutions of the system (\ref{Guco})-(\ref{Guco17}) above simplifies notably. In particular, from the first relation of \eqref{deltakappa}, one obtains the linear equation 
\begin{equation}\label{FGHJ}
0 = \kappa \partial_v  f + \partial_v^2f, 
\end{equation}
which has solution of the form
\begin{equation}
f(z ,\bar{z}, v) = T(z , \bar{z }) + e^{-\kappa  v}\ X(z , \bar{z }) \ .\label{Either}
\end{equation}

This generalizes the result of Ref. \cite{Donnay:2015abr}; see also \cite{Eling:2016xlx}. We will see below that, at the level of the asymptotic Killing vectors, (\ref{Either}) yields two (not mutually commuting) supertranslation currents associated to $T(z , \bar{z})$ and $X(z , \bar{z })$. 
The algebra \eqref{closure} now closes with
\begin{equation}
\label{algebrakfijo}
\badat{3}
&T_{12}=Y_1^A \pr_{A} T_2-Y_2^A\pr_{A} T_1,\\
&X_{12}=Y_1^A  \pr_{A}  X_2 -Y_2^A \pr_{A} X_1-\kappa(T_1 X_2-T_2 X_1),\\
&Y_{12}^A=Y_1^B  \pr_{B}  Y_2^A -Y_2^B  \pr_{B} Y_1^A.  
\eadat
\end{equation}

Let us represent the asymptotic Killing vector as $\chi=\chi(T,X,Y^{z},Y^{\bar{z}})$. By defining the Fourier modes, $T_{(m,n)}=\chi(z^m \bar z^n,0,0,0)$, $X_{(m,n)}=\chi(0,z^m \bar{z}^n,0,0)$, $Y_n=\chi(0,0,-z^{n+1},0)$ and $\bar Y_n=\chi(0,0,0,- \bar {z}^{n+1})$, we find
\begin{equation}\label{algebraa}
\badat{6}
&[ Y_m , Y_n ] = (m-n) Y_{m+n}, \\  
&[ \bar Y_m , \bar Y_n ] =  (m-n) \bar Y_{m+n},\\
&[ Y_k , T_{(m,n)} ] =  - m T_{(m+k,n)},\\
&[\bar Y_k , T_{(m,n)} ] =  - n T_{(m,n+k)},\\
&[ Y_k , X_{(m,n)} ] =  - m X_{(m+k,n)},\\
&[\bar Y_k , X_{(m,n)} ] =  - n X_{(m,n+k)},\\
&[ X_{(k,l)} ,T_{(m,n)}] = \kappa X_{(m+k,n+l)},
\eadat
\end{equation}
the remaining commutators being zero. As anticipated above, this algebra contains two sets of supertranslations currents, given by $T_{(m,n)}$ and $X_{(m,n)}$. Besides, it contains two sets of Virasoro (Witt) currents which are in semi-direct sum with the supertranslations. 

Being the zero-mode $T_{(0,0)}$ the Killing vector generating rigid translations in the advanced time direction $v$, and consequently suitable to be associated with the energy, it is worthwhile noticing that there is a large set of generators that commute with it. These are $Y_m$, $\bar{Y}_m$, and $T_{(m,n)}$. We may refer to these as the generators of the {\it soft hairs}. The generators $X_{(m,n)}$, in contrast, behave under the action of $T_{(0,0)}$ as an expansion; namely $[X_{(m,n)} , T_{(0,0)}] = \kappa X_{(m,n)}$. 

Notice that if we exclude from (\ref{algebraa}) the ideal generated by $X_{(m,n)}$, the remaining algebra is reminiscent of the four-dimensional extended Bondi-Metzner-Sachs algebra ($bms_4$) studied in Refs. \cite{Barnich3, Barnich2, Barnich4}, which also includes two copies of Virasoro algebra and supertranslations. However, it is worthwhile pointing out that both algebras are different, as it is suggested by the fact that the structure constants in the products $[Y_k , T_{(m,n)}]$ and $[\bar{Y}_k , T_{(m,n)}]$ in (\ref{algebraa}) differ from those of $bms_4$. In order to distinguish them, we will denote the subalgebra of (\ref{algebraa}) that does not include $X_{(m,n)}$ as $bms_4^{\mathcal H}$.

We will discuss in section \ref{extremalucho} the case of extremal black holes, for which $\kappa $ vanishes. This limit is singular in the sense that, remarkably enough, the algebra obtained in the case of extremal configurations does not coincide with the limit $\kappa \to 0$ in (\ref{algebraa}) but rather an algebra that corresponds to interchanging in (\ref{algebraa}) the roles played by $T_{(m,n)}$ and $X_{(m,n)}$, where the last line becomes $[ X_{(k,l)} ,T_{(m,n)}] = T_{(m+k,n+l)}$.

\subsubsection{Integrability}\label{integrabilitY}

The algebra of the asymptotic Killings vectors is (\ref{algebraa}). The question thus arises as to whether the charges associated to these Killing vectors, whose form was given in (\ref{QLaura}), satisfy an isomorphic algebra. In order to answer this, we first need to study the additional restrictions demanded by integrability. Apart from the isolated horizon condition $\kappa = const$, in four dimensions we must also require the fourth term in (\ref{QLaura}) to be integrable. 

If $\kappa $ is constant and in addition we assume $\partial_v \Omega = 0 $, then (\ref{QLaura}) can be integrated and it yields
\begin{equation}
Q_{[f,Y^A]}=\frac{1}{16\pi G} \int dz d\bar{z} \sqrt{\gamma}  ( 2T\kappa \Omega 
- Y^A \theta_A \Omega) + Q_0,
 \label{E3677}
\end{equation}
where, despite $f$ in general depends on $v$, such dependence does not appear in the charge; that is, there is no contribution of $X(z , \bar{z})$. $Q_0$ stands in (\ref{E3677}) as an arbitrary integration constant that corresponds to the charge of the geometry that is considered as a reference background\footnote{In Ref. \cite{Donnay:2015abr}, the value of $Q_0$ was chosen such that the reference geometry corresponds to that of zero horizon area.}. That is, (\ref{E3677}) is the expression of the charge associated to $T(z , \bar{z})$ found in Ref. \cite{Donnay:2015abr}; however, here we are reobtaining this result from a much general analysis, which in particular takes into account the possibility of $\partial_v f\neq 0$. Only the modes of the soft hairs $Y_m$, $\bar{Y}_m$, and $T_{(m,n)}$ contribute to the charges. As a consequence, when $\partial_v \Omega =0$ the algebra generated by the latter asymptotic Killing vectors coincides with the $bms_4^{\mathcal H}$ algebra \cite{Donnay:2015abr}.

A rather different scenario is that in which $\partial_v \Omega \neq 0$. Because of the fourth term in (\ref{QLaura}), this case requires the additional integrability condition
\begin{equation}\label{El1}
\Omega^{-1} {\partial_v \Omega} = {2}A(\Omega),
\end{equation}
with $A(\Omega)$ being a function of $\Omega $. 

Field equations are of help in the issue of solving this condition, as they yield relations between the metric functions and their derivatives. In fact, equation (\ref{vv}) can be written as
\begin{equation}
\partial^2_v \Omega^{1/2} = \kappa \partial_v\Omega^{1/2} , \label{El2}
\end{equation}
which, once combined with (\ref{El1}), has a solution
\begin{equation}
A(\Omega ) = \alpha \Omega^{-1/2} + \kappa, \label{El3}
\end{equation}
where $\alpha $ is an arbitrary constant. 

The next step is to check whether these conditions are preserved by the functional variations (\ref{deltakappa}); that is, whether the integrability condition is compatible with the asymptotic isometries. Let us consider the case $Y^A=0$ for simplicity. In this case, $\delta \Omega = f \partial_v\Omega $, and the variation of the functions $\Omega $ and $\partial_v\Omega$, together with (\ref{El3}) yield
\begin{equation}
\partial_vf  =0.
\end{equation}

Evaluating the expression of the charges, one eventually finds
\begin{equation}
Q_{[T,Y^A]} =\frac{1}{16\pi G} \int dz d\bar{z} \sqrt{\gamma}  ( 2T\kappa \Omega 
- Y^A \theta_A \Omega) + Q_0
. \label{E36777}
\end{equation}
That is, the functional form coincides with that of (\ref{E3677}). Nevertheless, it is worth emphasizing that while in (\ref{E36777}) $\Omega $ depends on time and $f$ does not, in (\ref{E3677}) function $\Omega $ was assumed not to depend on time and $f$ was in principle time dependent. Therefore, although the modes $X_{(m,n)}$ appear in the algebra of asymptotic Killing vectors, they do not contribute to the charges (\ref{E3677}), (\ref{E36777}) and, in this sense, seem to be pure gauge. The question arises as to how define charges in such a way that $X$ does contribute. We will discuss this issue below.

\subsection{Improved Dirac brackets} \label{labelDirac}

 There is a systematic way of constructing Dirac brackets in order to deal with cases in which, like in (\ref{QLaura}), the variation of the charges cannot be integrated without imposing ad hoc integrability conditions. This method has been developed in  \cite{Barnich4}  for surface charges and generalized for current algebras in \cite{Barnich6}.  
 
 Going back to expression (\ref{QLaura}), we follow \cite{Barnich4} and split the expression of the variation of the charge associated to a given Killing vector $\xi$ between its integrable part, denoted by $\delta Q^I_{\xi}[\Phi]$, and its non-integrable part, denoted by $\Theta_{\xi} [\Phi , \delta \Phi ]$, where $\Phi$ denotes the collection of all fields $\partial_v {\Omega}$, $\Omega$ and $\theta_A$. That is
\begin{equation}
\slashed{\delta} Q_{\xi}[\Phi ]= \delta Q_{\xi}^I[\Phi ] + \Theta_{\xi} [\Phi , \delta \Phi ] , \label{KLKL}
\end{equation}
where
\begin{equation}\label{CocoSilli}
Q^I_{\xi}[\Phi] =\frac{1}{16\pi G} \int dz d\bar{z} \sqrt{\gamma} \left( 2f\kappa \Omega + 2 \pr_v f  \Omega - 2 f \pr_{v} \Omega - Y^A \theta_A \Omega  \right) + Q_0
\end{equation}
and
\begin{equation}
\Theta_{\xi} [\Phi , \delta \Phi ] =\frac{1}{16\pi G} \int d z d\bar{z} \sqrt{\gamma} \ f \frac{\p_v \Omega}{\Omega} \delta \Omega   .
\end{equation}
The method presented in \cite{Barnich4} is the following: while in the integrable case one has
\begin{equation}
\label{bracketint}
\{ Q_{\xi_1}[\Phi ], Q_{\xi_2}[\Phi ] \}  = \delta_{\xi_2} Q_{\xi_1}[\Phi ] = -\delta_{\xi_1} Q_{\xi_2}[\Phi ]  ,
\end{equation}
in the case of non-integrable charges one can generalize this definition by considering the improved bracket
\begin{equation}\label{modifieddddd}
\{ Q^{I}_{\xi_1}[\Phi ], Q^{I}_{\xi_2}[\Phi ] \} ^* \equiv \delta_{\xi_2 } Q^I_{\xi_1}[\Phi] + \Theta_{\xi_2} [\Phi ,\delta_{\xi_1 } \Phi]  ,
\end{equation}
which only involves the integrable part of the charges on the left hand side and includes the non-integrable piece on the right hand side. After a lengthy computation\footnote{Where it is assumed that there are no obstructions to integrate by parts.}, one verifies that the right hand side of (\ref{modifieddddd}) can actually be gathered in the form (\ref{CocoSilli}). That is, the improved Dirac bracket (\ref{modifieddddd}) closes 
\begin{equation}
\{ Q^{I}_{\xi_1}[\Phi ], Q^{I}_{\xi_2}[\Phi ] \} ^*= Q^I_{\xi_{12}}[\Phi] ,
\end{equation}
with the Killing vectors $\xi _{12}$ given by (\ref{modified})-(\ref{modifieddd}). Therefore, (\ref{modifieddddd}) yields a representation of the algebra \eqref{algebrakfijo} including now the dependence on $X$. 

The charges $Q^I$ defined above satisfy the equation
\begin{equation}
\frac{d}{d v} Q^I_{\xi} [\Phi ] = - \Theta_{f=1,Y^A=0} [\Phi , \delta_\xi \Phi],
\end{equation}
which takes the form of an integrated continuity equation, the source being given by the non-integrable piece. This equation controls the non-conservation of the integrable part of the charges, which in general depends on time. This is analogous to what happens in the $bms_4$ case \cite{Barnich4}. 

\subsection{The zero-modes}

Now, having derived the charges and having discussed the integrability and conservation conditions, let us analyze their physical meaning by evaluating the integrated version of expression (\ref{QLaura}) on solutions whose interpretation is under control. Let us begin by considering the case of stationary black holes.

\subsubsection{Stationary black holes}

The Kerr metric written in the Eddington-Finkelstein coordinates is given by
\begin{equation}
\badat{2}
ds^2 =& \left( \frac{\Delta - \Xi }{\Sigma } -1\right) dv^2 + 2\ dv\ dr - \frac{2a(\Xi -\Delta )\sin^2\theta }{\Sigma } dv\ d\varphi - \\
&-2a\sin^2\theta \ dr\ d\phi +\Sigma d\theta^2 +\frac{(\Xi^2-a^2\Delta \sin^2\theta)\sin^2\theta}{\Sigma } d\varphi^2,
\eadat
\end{equation} 
where the functions $\Delta, \ \Xi,$ and $\Sigma $ are given by
\begin{equation}
\Delta (r)= r^2-2GM r+a^2 \ , \ \ \ \Xi (r)= r^2+a^2 \ , \ \ \ \Sigma (r)=r^2+a^2\cos^2\theta , \label{Kerrucho}
\end{equation}
where $M$ is the mass and $a$ is the angular momentum per unit of mass. The outer horizon of the Kerr black hole is located at $r_+=GM+\sqrt{G^2M^2-a^2}$.

Kerr metric can be written in the form \eqref{eql:null333} and \eqref{eql:null4}. The explicit change of coordinates can be found, for instance, in \cite{Booth:2012xm}. In these coordinates, the metric reads 
\begin{equation}
g_{\rho v }=1  , \ \ g_{\rho \varphi }=0 , \ \ g_{\rho \theta }=0 , \ \ g_{\rho \rho }=0,
\end{equation}
together with
\begin{equation}
\kappa = - \frac{\Delta '(r_+ )}{2\Xi (r_+ )},
\end{equation}
where $\Delta '(r_+)=2(r_+ -GM)= (r_+^2 -a^2)/r_+ $,
\begin{equation}
\theta_{ \theta} =\frac{2a^2\sin\theta \cos\theta }{\Sigma (r_+ )}, \ \ \ 
\theta_{\varphi } = -\left( \frac{a\Delta '(r_+)\sin^2\theta }{\Sigma (r_+)} + \frac{2ar_+\Xi (r_+)\sin^2\theta }{\Sigma^2(r_+)} \right),
\end{equation}
and 
\begin{equation}\label{LosThetas}
\Omega_{\theta \theta } =  \Sigma (r_+) , \ \ \ 
\Omega_{\theta \varphi } =  0, \ \ \ 
\Omega_{\varphi \varphi } =   \frac{\Xi^2 (r_+)\sin^2\theta }{\Sigma (r_+)} .
\end{equation}
Notice that, since the Hawking temperature of the Kerr black hole is given by 
\begin{equation}
T=\frac{1}{4\pi }\frac{\Delta '(r_+)}{\Xi (r_+ )}, 
\end{equation}
then one finds $g^{(1)}_{vv}= -{\Delta ' (r_+)}/{\Xi (r_+ )}= -2\kappa $, in accordance with the near horizon expansion \eqref{eql:null4} and with the identification of the function $\kappa $ with the surface gravity. Introducing a {field-dependent} change of coordinates
\begin{equation}
z= e^{i\varphi }\mu(\theta),\quad \mu(\theta)=\cot (\theta /2)e^{-\frac{a^2}{r_+^2+a^2}\cos(\theta)},
\end{equation}
one manages to write the metric of the horizon in the conformal form
\begin{align}
ds^2_H=\Omega \gamma_{AB} dx^A dx^B,
\end{align}
with
\begin{align}
\Omega=\frac{(r_+^2+a^2)^2}{r_+^2+a^2 \cos(\theta)^2}\left(\cos(\theta/2)^2 e^{-\frac{a^2}{r_+^2+a^2} \cos(\theta)}+\sin(\theta/2)^2e^{\frac{a^2}{r_+^2+a^2}\cos(\theta)}\right)^2.
\end{align}

Then, evaluating the charges ${\mathcal T}_{(m,n)}$ for the Kerr metric, integrating between two black hole configurations $(A)$ and $(B)$ at fixed temperature $\kappa / (2\pi )$, one finds
\begin{equation} \label{E47}
{\mathcal T}^{(A)}_{(0,0)} - {\mathcal T}^{(B)}_{(0,0)} = T\ \frac{\Delta {\mathcal A}}{4G} \ , \ \ \ \ T=\frac{\kappa }{2\pi } ,
\end{equation}
where $\Delta {\mathcal A}$ is the difference between the area corresponding to the configurations $(A)$ and $(B)$. This is independent of $Q_0$. That is, the charge
${\mathcal T}_{(0,0)}$ gives actually the Bekenstein-Hawking entropy. 

On the other hand, the charges associated to superrotations read\footnote{Here we set $Q_0=0$.}
\begin{equation}\label{E48}
\mathcal{Y}_n=\frac{1}{16 \pi G} \int{dz d\bar{z}} \sqrt{\gamma} \Omega z^{n+1}\theta_{z},
\end{equation}
which, using the coordinates definition above, can be written as
\begin{equation} \label{E48bis}
\mathcal{Y}_n=i \frac{Ma}{2} \delta_{n,0},
\end{equation}
and, analogously,
\begin{equation} \label{E48bisbis}
\mathcal{\bar Y}_n=-i \frac{Ma}{2} \delta_{n,0}.
\end{equation}
Here, we have used that $\theta_A dx^A=\theta_\theta d \theta + \theta_\varphi d \varphi=\theta_z d z + \theta_{\bar{z}} d \bar{z}$ and (\ref{LosThetas}), which yields
\begin{equation}
z \theta_{z}=\frac{1}{2} \left( \frac{\mu}{\mu'} \theta_\theta -i \theta_{\varphi}\right), \quad \bar{z} \theta_{\bar{z}}=\frac{1}{2} \left( \frac{\mu}{\mu'} \theta_\theta +i \theta_{\varphi}\right).
\end{equation}
Expressions (\ref{E47}), (\ref{E48bis}), and (\ref{E48bisbis}) provide us with a clear physical interpretation of the charges. In the case of stationary black holes the only non-vanishing charges are the zero-modes, corresponding to the Wald entropy and the angular momentum \cite{Donnay:2015abr}. Let us see now whether such an interpretation also holds in other examples.

\subsubsection{Extremal limit and NHEK geometry} \label{extremalucho}

Let us here consider the extremal limit $a^2 \to (GM)^2$. In this case, $\kappa $ vanishes and this is why this limit requires a separated analysis. 

In the coordinates we used above to describe the Kerr geometry, the extremal case corresponds to the Near-Horizon-Extremal-Kerr (NHEK) geometry \cite{BardeenHorowitz}, which is relevant for Kerr/CFT \cite{KerrCFT}. Explicitly, when $a=GM$, then $r_+=a$ and one finds $\kappa = 0$ together with
\begin{equation}
\theta_{ \theta} =\frac{2\sin\theta \cos\theta }{1+\cos^2\theta }, \ \ \ 
\theta_{\varphi } = - \frac{4\sin^2\theta }{(1+\cos^2\theta )^2} ,
\end{equation}
and
\begin{equation}\label{LosThetasEXTRE}
\Omega_{\theta \theta } =  a^2(1+\cos^2\theta ) , \ \ \ 
\Omega_{\theta \varphi } =  0, \ \ \ 
\Omega_{\varphi \varphi } =   \frac{4a^2\sin^2\theta }{1+\cos^2\theta }.
\end{equation}
To analyze this special case, one has to be careful because the expression of the charges change: while the derivation of the variation $\slashed{\delta} Q$ did not assume $\kappa \neq 0$, it turns out that the integrability condition and the solution to the differential equation (\ref{FGHJ}) get modified when $\kappa $ vanishes. In particular, one finds 
\begin{equation}
f=T(z,\bar{z})+v\ X(z,\bar{z}) , \label{fExtremal}
\end{equation} 
with $z$ now being $z=e^{i\varphi -\cos (\theta)/2} \cot (\theta/2)$, while the integrability condition demands $\partial_v \Omega =0$. In addition, the absence of the first term in (\ref{QLaura}) when $\kappa =0$ produces a contribution of function $X$ in the charge $Q$.  This yields the following form for the charge in the extremal case
\begin{equation}\label{notegusta64}
Q_{[f,Y^A]} =\frac{1}{16\pi G} \int dz d\bar{z} \sqrt{\gamma}  \Bigg( 2X \Omega 
- Y^A \theta_A \Omega
\Bigg) +Q_0 .
\end{equation}

Again, if evaluating this charge in the stationary solution, where only the zero-modes of the charges contribute, one finds that the charge associated to $X_{(0,0)}$ yields
\begin{equation} \label{E47Extremal}
{\mathcal X}^{(A)}_{(0,0)} - {\mathcal X}^{(B)}_{(0,0)} = T_L \ \frac{\Delta {\mathcal A}}{4G} \ , \ \ \ \ T_L = \frac{1}{2\pi } 
\end{equation}
where, as in (\ref{E47}), $\Delta {\mathcal A}$ is the difference between the area corresponding to the configurations $(A)$ and $(B)$. However, here the zero mode $X_{(0,0)}$ is identified with the temperature that appears in the Kerr/CFT analysis of the extremal Frolov-Thorne vacuum \cite{FrolovThorne, KerrCFT}, $T_L$.

This shows that, even when the extremal limit is somehow singular and the algebra of charges and integrability conditions are modified\footnote{See in particular the difference between (\ref{notegusta64}) and (\ref{E36777}).}, our computation of the charges in the near horizon region yields the correct answer for the black hole entropy.

\subsubsection{Rindler horizons}

Other horizons in four-dimensions we can analyze are those experienced by an uniformly accelerating observer in flat space, namely the Rindler horizon. In fact, Rindler spacetime in the region close to the wedge can also be gathered in the form \eqref{eql:null3} and \eqref{eql:null4}. More precisely, it reads
\begin{equation}
ds^2 = -2\kappa \ \rho \ dv^2 + 2\ dv\ d\rho + 2\ dz d\bar{z} ,
\end{equation}
where $T=\kappa /(2\pi) = a/(2\pi )$ is now the Unruh temperature, with $a$ being the acceleration of the observer. This corresponds to \eqref{conformalhorizon} with $\Omega = (1+z \bar{z})^2/2 = 1/\sqrt{\gamma }$. Computing the zero-mode of the charge, one finds
\begin{equation}\label{Laflamme}
{\mathcal T}_{(0,0)} / {\mathcal A} = T \ \frac{1}{4G} \ , \ \ \ \ T = \frac{a}{2\pi },
\end{equation}
where we now choose to write the charge per unit of area because, strictly speaking, the entropy of Rindler spacetime is infinite and only the entropy per unit of area makes sense. Here, we define the charge ${\mathcal T}_{(0,0)}$ with respect to the solution with $a=0$, which is part of the solution space, setting  $Q_0=0$.

We notice that (\ref{Laflamme}) exactly reproduces the Laflamme result for the Rindler density entropy \cite{Laflamme}. This is again consistent with the interpretation of the zero-mode of the charge to be the Wald entropy.

\section{Three-dimensional horizons}

As mentioned in subsection \ref{conno}, the problem of the charge integrability notably simplifies in three spacetime dimensions. This results in an enhancement of the charge algebra. More precisely, unlike what happens in four dimensions, in three dimensions considering time dependence in the function $f$ that defines the component $\chi ^v$ of the asymptotic Killing vector makes the function $X$ to appear in the expression for the charge without resorting to the improvement of the Dirac bracket discussed in section \ref{labelDirac}. As a consequence, the charge algebra occurs to contain two sets of supertranslations, making the parallel with what happens at the level of the asymptotic Killing vector algebra. This generalizes the result of Ref. \cite{Donnay:2015abr}, where only one set of supertranslations was found. Another advantage of performing the analysis in three dimensions is that the field equations can be solved in a closed way, allowing us to present a family of exact solutions that realizes the proposed near boundary conditions explicitly. We will begin the study of three-dimensional horizons by defining the boundary conditions in the next subsection.

\subsection{Boundary conditions and charges}

In the near horizon region, the metric takes the form 
\begin{equation}
\label{BChorizon}
ds^2=-2\kappa \rho \ dv^2 + 2d\rho \ dv +2\theta \rho \ dv\ d\phi +(\gamma^2+\lambda \rho) d\phi^2 + \Delta g_{i j} dx^{i} dx^{j},
\end{equation}
where $\kappa$, $\theta$, $\gamma$, $\lambda$ are functions that depend explicitly on $v$ and $\phi$ and the components of $ \Delta g_{i j}$ are $O(\rho^2)$ for $i,j \in \{v,\phi\}$. This is the analogous of the boundary conditions (\ref{formA}) considered in the four-dimensional case. 

This set of asymptotic conditions is preserved by the following transformations
\be
\badat{3}
  \label{eq:3}
&\chi^{v}=f(v,\phi), \\ 
&\chi^{\rho}=-\pr_{v} f \rho +\pr_{\phi} f  \frac{\theta}{2\gamma^2} \rho^2 + O(\rho^3),\\
&\chi^{\phi}=Y(\phi)-\pr_{\phi} f \frac{\rho}{\gamma^2}+ \pr_{\phi} f \frac{\lambda}{2\gamma^4}  \rho^2 + O(\rho^3).
\eadat
\en
Note that the component $\chi^{\rho}$ may contain a $O(1)$ term $Z(v,\phi)$. However, as we have shown in section \ref{s4D}, such term is set to zero provided function $Y$ is field independent. 
Under these transformations, the fields appearing in the metric transform as
\be
\badat{4}
&\delta_{\chi} \kappa =Y \pr_{\phi} \kappa + \pr_v (\kappa f) + \pr_v^2 f, \label{deltakappa3d} \\ 
&\delta_{\chi}  \gamma = \pr_{\phi}(Y \gamma) + f \pr_v \gamma,\\
&\delta_{\chi}  \theta=\pr_{\phi}(Y \theta) + f \pr_v \theta -2\kappa \pr_{\phi} f -2\pr_v \pr_\phi f + 2 \pr_\phi f \frac{\pr_v \gamma}{\gamma},\\
&\delta_{\chi}  \lambda= Y\pr_\phi \lambda+ 2\lambda \pr_\phi Y +2\theta \pr_{\phi} f -2\pr^2_\phi f + 2 \pr_\phi f \frac{\pr_\phi \gamma}{\gamma}+f \pr_v \lambda- \lambda \pr_v f.
\eadat
\en
By introducing the modified Lie brackets \cite{Barnich3}
\begin{equation}
\label{modifed}
  [\chi_1,\chi_2]=\mathcal{L}_{\chi_1}\chi_2-\delta_{\chi_1} \chi_2+\delta_{\chi_2} \chi_1 ,
\end{equation}
one finds that, as it happens in four dimensions, the algebra of these vectors closes, 
\begin{equation}
  [\chi(f_1,Y_1),\chi(f_2,Y_2)]=\chi(f_{12},Y_{12}),
\end{equation}
where now
\begin{equation}
\begin{split}
f_{12}=f_1 \pr_{v} f_2-f_2\pr_{v} f_1+Y_1 \pr_{\phi} f_2-Y_2\pr_{\phi} f_1,\quad Y_{12}=Y_1  \pr_{\phi}  Y_2 -Y_2  \pr_{\phi} Y_1.  
\end{split}
\end{equation}
These symmetries yield the following expression for the variation of the charges
\begin{equation}
\slashed{\delta} Q_{[f,Y]} =\frac{1}{16\pi G} \int^{2\pi}_{0} d\phi \left( 2f\kappa \delta \gamma + 2 \pr_v f \delta \gamma - 2 f \delta \pr_{v} \gamma - Y\delta(\theta \gamma)  \right),
\label{QLaura2}
\end{equation}
where now, as mentioned in section 2, there is no analog to the fourth term in (\ref{QLaura}). This means that demanding $\kappa $ to be constant would be sufficient to guarantee integrability of $Q_{[f,Y]}$.

\subsection{Field equations}

The Einstein's field equations (with or without cosmological constant) can be solved perturbatively order by order in the $O (\rho ^n )$ expansion. By considering spacetimes \eqref{BChorizon}, we find that order $O(\rho^0)$ of Einstein's equation components $(v,v)$ and $(v,\phi)$ impose
\begin{equation}
\label{Einstein3d}
 \partial_v(\gamma \theta)+2\gamma \partial_\phi \kappa=0, \quad \pr_v^2 \gamma =\kappa \pr_v\gamma. 
 \end{equation}
These relations are sufficient to show conservation and closure of the algebra. However,  as we will see in section \eqref{subsection34},  
we can completely solve Einstein equations, finding an exact solution depending on three arbitrary functions of $\phi$.  

\subsection{Fixed temperature configurations}

Again, we will impose that $\kappa$ is fixed. Replacing this in equation \eqref{deltakappa3d}, we find
\begin{equation}
f(v,\phi)=T(\phi)+e^{-\kappa v} X(\phi).
\label{kcst}
\end{equation}
The algebra turns out to be
\begin{equation}
  [\chi(T_1,X_1,Y_1),\chi(T_2,X_2,Y_2)]=\chi(T_{12},X_{12},Y_{12}),
\end{equation}
where
\begin{equation}
\begin{split}
T_{12}&=Y_1 \pr_{\phi} T_2-Y_2\pr_{\phi} T_1,\\ 
X_{12}&=Y_1 \pr_{\phi} X_2-Y_2\pr_{\phi} X_1-\kappa(T_1 X_2 - T_2 X_1),\\ 
Y_{12}&=Y_1  \pr_{\phi}  Y_2 -Y_2  \pr_{\phi} Y_1.
\end{split}
\end{equation}

By defining the Fourier modes, $T_n=\chi(e^{in\phi},0,0)$, $X_n=\chi(0,e^{in\phi},0)$ and $Y_n=\chi(0,0,e^{in\phi})$, we find
\begin{equation}\label{algebraX3d}
\begin{split}
i[ Y_m , Y_n ] &= (m-n) Y_{m+n}, \\  
i[ Y_m , T_n ] &=  - n T_{m+n},\\
i[ Y_m , X_n ] &=  - n X_{m+n},\\
i[ T_m , X_n ] &= -i\kappa X_{m+n}.
\end{split}
\end{equation}
In this case, the surface charges \eqref{QLaura2} can be integrated as 
\begin{equation}
\label{Qconserved}
 Q_{[T,X,Y]}=\frac{1}{16\pi G} \int^{2\pi}_{0} d\phi \left( 2T(\kappa  \gamma-\pr_{v} \gamma) - 2 X e^{-\kappa v} \pr_v \gamma- Y \theta \gamma  \right) +Q_0,
\end{equation}
where $Q_0$ is a constant without variation (that has been considered to be zero in \cite{Donnay:2015abr}).
The charge \eqref{Qconserved} is conserved, $\partial_v Q=0$, by virtue of the relations imposed by the field equations \eqref{Einstein3d}.
Furthermore, since the charges are integrable we can prove that
\begin{equation}
\label{3d-alg}
 \{ Q_{[T_1,X_1,Y_1]},Q_{[T_2,X_2,Y_2]}\} =Q_{[T_{12},X_{12},Y_{12}]} ,
\end{equation}
by virtue of \eqref{bracketint}.

In conclusion, the charges are integrable, conserved and form a representation of \eqref{algebraX3d}. This result generalizes the one found in 
\cite{Donnay:2015abr}: we see from (\ref{algebraX3d}) that a new set of supertranslations associated to the function $X(\phi)$ appears. The subalgebra generated by $X_n$ is an ideal of the charge algebra that, unlike the other infinite charges, do not commute with $T_0$.

\subsection{Exact solution} \label{subsection34}

In three-dimensional general relativity there is a relevant solution to which all this discussion concerns: the Ba\~{n}ados-Teitelboim-Zanelli (BTZ) black hole \cite{BTZ}, which is a solution of the theory in presence of a negative cosmological constant $\Lambda <0$. The purpose of this subsection is to present a family of exact solutions of Einstein's field equations that, while gathering the BTZ black hole as a particular case, happens to realize the boundary conditions (\ref{BChorizon}) explicitly. Assuming $\kappa \neq 0$, the explicit form of the metric reads
\begin{equation}\label{soportame83}
\badat{4}
g_{vv}&=-2\kappa\rho +  \left( \frac{\theta^2}{4 \gamma^2} -\frac{1}{\ell^2}\right)\rho^2    , \\ 
g_{v \phi}&=\theta \rho + \frac{\theta \lambda}{4 \gamma^2}\rho^2,\\
g_{\phi \phi}&=\left( \gamma + \frac{\lambda}{2 \gamma}\rho\right)^2,\\
g_{\rho v}&=1, \\ 
g_{\rho \rho}&=g_{\rho \phi}=0,
\eadat
\end{equation}
where $\gamma$ and $\theta$ satisfy equations \eqref{Einstein3d} and $\lambda$ is obtained from
\begin{equation}
\partial_v {\lambda}+\left(\kappa -\frac{\partial_v {\gamma}}{\gamma}\right) \lambda=\partial_{\phi }\theta-\frac{1}{2} \theta^2+\frac{2}{\ell^2} \gamma^2-\theta \frac{\partial_{\phi }\gamma }{\gamma}.
\end{equation}

In the particular case $\partial_v \lambda = \partial_v \gamma =0$ this yields the solution found in Ref. \cite{Donnay:2015abr}. In the general case, the solution reads
\begin{equation}\label{soportame84}
\badat{4}
 \gamma&=\gamma_0+e^{\kappa v}\eta ,\quad \theta=\frac{\cJ}{\gamma},\\
  \kappa \lambda&=\frac{1}{\ell^2}\left(\gamma^2+\gamma \gamma_0 \right)+e^{-\kappa v}\left(C \gamma +\frac{\cJ^2}{4 \eta \gamma}+\frac{\cJ}{\gamma}\pr_{\phi}\left( \frac{\gamma_0}{\eta}\right)-\eta\partial_\phi \left(\frac{\cJ}{\eta^2} \right)     \right)
\eadat
\end{equation}
while  $\gamma_0$, $\cJ$, $\eta$ and $C$ are arbitrary functions of $\phi$ $(\eta\neq 0)$; $C$ does not ultimately appear in the charges. Function $C$ can be fixed by initial conditions at $v=0$ from equations (\ref{soportame84}) as a function of $\theta_{(v=0)}$, $\gamma_0 $ and $\eta $. This family of solutions generalizes the one presented in \cite{Donnay:2015abr}. 
 
Evaluated on these geometries, the charges (\ref{Qconserved}) take the form
\begin{equation}
Q_{[T,X,Y]} = \frac{1}{16\pi G} \int^{2\pi }_0 d\phi (2T\kappa \gamma_0 -2X\kappa \eta -Y \theta \gamma ) +Q_0 ,
\end{equation}
with the first term on the right hand side being the one that gives the entropy in the case of stationary solutions. 

\subsection{Extremal black holes}

Let us now study the limit in which $\kappa $ tends to zero. This is what happens in the case of extremal black holes. In such case, the boundary conditions will be similar to \eqref{BChorizon}, but with $g_{vv}$ given by
\begin{equation}
\label{extremal}
g_{vv}=L(v,\phi) \rho^2+O(\rho^3).
\end{equation}
Since the function $L(v,\phi)$ is allowed to vary arbitrarily, the asymptotic Killing vector analysis results to be the same and one ends up with the general form \eqref{eq:3}, the only difference being that now the equation for $f$ coming from \eqref{deltakappa3d} yields
\begin{equation}
f(v,\phi ) = T(\phi ) + v\ X(\phi ).
\end{equation}

Again, the algebra closes under the modified Lie bracket
\begin{equation}
  [\chi(T_1,X_1,Y_1),\chi(T_2,X_2,Y_2)]=\chi(T_{12},X_{12},Y_{12}),
\end{equation}
where
\begin{equation}
\begin{split}
T_{12}&=T_1 X_2 - T_2 X_1+Y_1 \pr_{\phi} T_2-Y_2\pr_{\phi} T_1,\\ 
X_{12}&=Y_1 \pr_{\phi} X_2-Y_2\pr_{\phi} X_1,\\ 
Y_{12}&=Y_1  \pr_{\phi}  Y_2 -Y_2  \pr_{\phi} Y_1.
\end{split}
\end{equation}

By defining the Fourier modes, $T_n=\chi(e^{in\phi},0,0)$, $X_n=\chi(0,e^{in\phi},0)$ and $Y_n=\chi(0,0,e^{in\phi})$, we find
\begin{equation}\label{algebraext}
\begin{split}
i[ Y_m , Y_n ] &= (m-n) Y_{m+n}, \\  
i[ Y_m , T_n ] &=  - n T_{m+n},\\
i[ Y_m , X_n ] &=  - n X_{m+n},\\
[ T_m , X_n ] &= T_{m+n}.
\end{split}
\end{equation}

Remarkably, this algebra is not the $\kappa \gt 0$ limit of algebra (\ref{algebraX3d}). On the one hand, the bracket between supertranslation charges $T_n$ and $X_n$ is not found to vanish; on the other hand, there is an interchange between these two sets of charges. Nevertheless, a single Virasoro algebra in semi-direct sum with two affine currents is still the algebra that appears in the near horizon region of the extremal solutions.

The solution in the case $\kappa =0$ is similar to (\ref{soportame83})-(\ref{soportame84}) but replacing functions $\gamma $ and $\lambda $ by
\begin{equation}
\badat{4}
\gamma &= \gamma _0 + v \ F,\\
\lambda&=\frac{1}{\ell^2} v \left(\gamma^2+\gamma \gamma_0 \right)+C \gamma +\frac{\cJ^2}{4 \eta \gamma}+\frac{\cJ}{\gamma}\pr_{\phi}\left( \frac{\gamma_0}{\eta}\right)-\eta\partial_\phi \left(\frac{\cJ}{\eta^2} \right)    
\eadat
\end{equation}
This eventually yields the charge
\begin{equation}
Q_{[T,X,Y]} = \frac{1}{16\pi G} \int^{2\pi }_0 d\phi (2X \gamma_0 -2T \eta -Y \theta \gamma ) +Q_0 ,
\end{equation}
where we see again the interchange between the role played by $X$ and by $T$ with respect to the non-extremal case. In the extremal case, it is the charge associated to zero mode of $X$ the one that gives the entropy in the case of stationary configurations.

\subsection{Other boundary conditions}

\subsubsection{Early works}

Before concluding, we would like to discuss the relation between our analysis of the near horizon symmetries and other results in the literature. In fact, the analysis of asymptotic isometries in the near horizon region has been studied long time ago; see for instance \cite{Carlip1, Koga, Hotta:2002mq, Hotta:2000gx} and references therein and thereof. In \cite{Hotta:2002mq}, a different set of boundary conditions is considered in the three-dimensional case\footnote{See, for instance, the components $g_{\rho v}$ in Eq. (40) therein; see also the differences in dependences allowed in the asymptotic Killing vectors (42)-(44).}, yielding the same algebra for the charges. The analysis of the charges in \cite{Hotta:2002mq}, however, differs from the one considered here. In our case, it is not necessary to introduce any constant to be ulteriorly set to the Planck scale, but our expressions for the charges smoothly follows from the covariant approach \cite{Barnich:2001jy, Barnich:2007bf}. From our definition of the charges, on the other hand, it turns out to be clear which are the assumptions required for integrability, and why the condition of $\kappa $ being fixed is important to that end. 

In Ref. \cite{Hotta:2000gx} the asymptotic isometry of non-rotating black hole in four dimensions were studied within a formalism similar to that of \cite{Hotta:2002mq}. In that case, the algebra of charges obtained is different from the one obtained in section \ref{conno} herein, and the reason is that the boundary conditions are different. In particular, our charges also reproduce the angular momentum of spinning Kerr black holes.

\subsubsection{Recent works}

More recently, after Ref. \cite{Hawking} appeared, there has been a renewed interest in the problem of asymptotic symmetries close to the horizon of non-extremal black holes. This led to consider different sets of boundary conditions. For instance, in Ref. \cite{Afshar:2016wfy} an infinite-dimensional algebra different from ours has been derived. This follows from considering a different set asymptotic expansion at the horizon. Unlike the algebra derived here (and in Ref. \cite{Donnay:2015abr}), the algebra found in \cite{Afshar:2016wfy} (and in Ref. \cite{Afshar:2016uax}) consists of two copies of the affine extension of Heisenberg algebra; that is, two mutually commuting copies of the affine Kac-Moody extension of $u(1)$. These are centrally extended supertranslations. 

The boundary conditions considered in \cite{Afshar:2016wfy} are defined by considering metrics of the form
\begin{equation}
ds^2=-2a \rho \ell f dv^2+2\ell dv\ d\rho -2\omega a^{-1}d\varphi \ d\rho + 2\omega \rho f dv \ d\varphi+(\gamma^2+2\rho a^{-1}f(\gamma^2-\omega ^2)/\ell))d\varphi^2, \label{Uhhhh90}
\end{equation}
where $f=1+(2a\ell)^{-1}\rho$. By performing the supertranslation change of coordinates defined by 
\begin{equation}
v\to v + T(\varphi ) \ , \ \ \ \text{with} \ \ \ T(\varphi )= \frac{1}{a\ell}\int^{\varphi } \omega(\phi )d\phi 
\end{equation}
one can bring metric (\ref{Uhhhh90}) into the form (\ref{BChorizon}), in the case $\theta =0$. However, the transformations considered in \cite{Afshar:2016wfy} are not of that class, and this is the reason why the algebra of asymptotic isometries and that of the associated charges are different from ours. It is worthwhile mentioning that an inspection to the origin of the two centrally extended supertranslations found in \cite{Afshar:2016wfy} is different from the origin of the two set of supertranslations appearing in (\ref{algebraX3d}).

\section{Conclusions}

In this paper we have shown that the near horizon geometry of non-extremal black holes exhibits an infinite-dimensional symmetry that extends supertranslations. In four spacetime dimensions, the full symmetry algebra has been shown to include two sets of supertranslations in semi-direct sum with two mutually commuting copies of Virasoro algebras. This is given in (\ref{algebraa}). This extends the analysis of Ref. \cite{Donnay:2015abr}, in particular by taking into account time-dependent configurations. This resulted in the enhancement of the symmetry algebra. We have discussed the proper definition of the surface charges associated to these symmetries and studied their integrability properties. With the adequate definition of the Dirac brackets, the charges were shown to close the same algebra as the one of the asymptotic Killing vectors that preserve the boundary conditions at the horizon. In the case of stationary black holes, the only charges that do not vanish turn out to be the zero-modes, one of which provides the entropy of the black hole. We also studied the extremal limit, in which the zero-mode of the charges also reproduces the entropy. 

Some question remain still open: First, it remains to be understood the physical meaning of the additional charges $\cY_m$, $\bar{\cY}_m$, and $\cT_{(m,n)}$ with $m,n\neq 0$ and the interplay with the supertranslation (and superrotation) hairs. Also in the case of the zero-modes $m=n=0$, the geometric interpretation of the charges for time-dependent configurations requires further study. It would be worthwhile investigating the relation between the analysis of the extremal limit carried out here and the standard setup of the Kerr/CFT correspondence. Last, it would be desirable to explore different sets of boundary conditions at the horizon and see how different choices lead to different infinite-dimensional symmetry algebras. 

\section*{Acknowledgments}

We thank Daniel Grumiller for interesting discussions. L.D. is a FRIA Research Fellow of the FNRS Belgium. G.G. is partially supported by CONICET through the research grant PIP 0595/13. H.G. is partially supported by the ERC Advanced Grant SyDuGraM, by FNRS-Belgium (convention FRFC PDR T.1025.14 and convention IISN 4.4503.15) and by the ``Communaut\'e Francaise de Belgique'' through the ARC program. M.P. is supported by FONDECYT/Chile grants 7912010045 and 11130083.


\newpage{\pagestyle{empty}\cleardoublepage}

\end{document}